\documentclass[graybox]{svmult}

\usepackage{mathptmx}       
\usepackage{helvet}         
\usepackage{courier}        
\usepackage{type1cm}        
\usepackage{makeidx}         
\usepackage{graphicx}        
\usepackage{multicol}        
\usepackage[bottom]{footmisc}

\makeindex             

\begin{document}

\title*{Pulsating B and Be stars in the Magellanic Clouds}
\author{P.~D.~Diago$^{1}$, J.~Guti\'{e}rrez-Soto$^{1,2}$, J.~Fabregat$^{1,2}$, C.~Martayan$^{2,3}$ and J.~Suso$^{1}$}
\authorrunning{P.~D.~Diago et al.}
\institute{$^{1}$ \at Observatori Astron\`{o}mic de la Universitat de Val\`{e}ncia, Ed. Instituts d'Investigaci\'{o}, Pol\'{i}gon La Coma, 46980 Paterna, Val\`{e}ncia, Spain. \email{Pascual.Diago@uv.es}
\and
$^{2}$ \at GEPI, Observatoire de Paris, CNRS, Universit\'{e} Paris Diderot, Place Jules Janssen 92195 Meudon Cedex, France.
\and
$^{3}$ \at Royal Observatory of Belgium, 3 Avenue Circulaire, B-1180 Brussels, Belgium.
}
%
%
\maketitle

\abstract*{Stellar pulsations in main-sequence B-type stars are driven by the $\kappa$-mechanism due to the Fe-group opacity bump. The current models do not predict the presence of instability strips in the B spectral domain at very low metallicities. As the metallicity of the Magellanic Clouds (MC) has been measured to be around $Z=0.002$ for the Small Magellanic Cloud (SMC) and $Z=0.007$ for the Large Magellanic Cloud (LMC), they constitute a very suitable objects to test these predictions.\newline\indent
The aim of this work is to investigate the existence of B-type pulsators at low metallicities, searching for short-term periodic variability in a large sample of B and Be stars from the MC with accurately determined fundamental astrophysical parameters.
}

\abstract{Stellar pulsations in main-sequence B-type stars are driven by the $\kappa$-mechanism due to the Fe-group opacity bump. The current models do not predict the presence of instability strips in the B spectral domain at very low metallicities. As the metallicity of the Magellanic Clouds (MC) has been measured to be around $Z=0.002$ for the Small Magellanic Cloud (SMC) and $Z=0.007$ for the Large Magellanic Cloud (LMC), they constitute a very suitable objects to test these predictions.\newline\indent
The aim of this work is to investigate the existence of B-type pulsators at low metallicities, searching for short-term periodic variability in a large sample of B and Be stars from the MC with accurately determined fundamental astrophysical parameters.
}

\section{Introduction}
\label{sec:1}

A significant fraction of main-sequence B-type stars are known to be variable. The whole main-sequence in the B spectral range is populated by stars belonging to two well characterised classes of pulsating variables: the $\beta$ Cephei stars (
\cite{stankov2005}) and the Slowly Pulsating B stars (SPB, 
\cite{decat2002}). Both types of stars pulsate due to the $\kappa$-mechanism acting in the partial ionization zones of iron-group elements. $\beta$ Cephei stars do pulsate in low-order p- and g-modes with periods similar to the fundamental radial mode. SPB stars are high-radial order g-mode pulsators with periods longer than the fundamental radial one.

Be stars are defined as non-supergiant B star whose spectrum has or had at some time one or more Balmer lines in emission. Classical Be stars are physically understood as rapidly rotating B-type stars with line emission arising from a circumstellar disk in the equatorial plane, made by matter ejected from the stellar photosphere by mechanisms not yet understood (see 
\cite{porter2003} for a recent review). Be stars show two different types of photometric variability with different origin and time-scale: (i) long-term variability due to variations in size and density of the circumstellar envelope. Variations are irregular and sometimes quasi-periodic, with time scales of weeks to years. (ii) short-term periodic variability, with time-scales from 0.1 to 3 days, attributed to the presence of non-radial pulsations. As pulsating Be stars occupy the same region of the HR diagram that $\beta$ Cephei and SPB stars, it is generally assumed that pulsations have the same origin, i.e., p and/or g-mode pulsations driven by the $\kappa$-mechanism associated to the Fe bump.

The $\kappa$-mechanism in $\beta$ Cephei and SPB stars has an important dependence on the abundance of iron-group elements, and hence the respective instability strips have a great dependence on the metallicity of the stellar environment. 
\cite{pamyatnykh1999} showed that the $\beta$ Cephei and SPB instability strips practically vanish at $Z < 0.01$ and $Z < 0.006$, respectively. The metallicity of the Magellanic Clouds (MC) has been measured to be around $Z= 0.002$ for the Small Magellanic Cloud (SMC) and $Z= 0.007$ for the Large Magellanic Cloud (LMC) (see 
\cite{maeder1999} and references therein). Therefore, it is  expected to find a very low occurrence of $\beta$ Cephei or SPB pulsators in the LMC and no pulsator types in the SMC.

\section{The data sample and the frequency analysis}
\label{sec:2}
Our research has been focused in a sample of more than 150 stars for the LMC and more than 300 stars for the SMC for which 
\cite{martayan2006,martayan2007} obtained medium resolution spectroscopy with FLAMES instrument at ESO/VLT. In these papers, the authors provided accurately fundamental astrophysical parameters and  this allows to place the periodic variables in the HR diagrams in order to map the regions of pulsational instability for the low-metallicity environments as the Magellanic Clouds. All the photometric time series had been retrieved from the MACHO project (
\cite{alcock1999}), because the time span is large enough to provide a very high frequency resolution in the spectral analysis and allows to distinguish between very close frequencies. The complete results of the frequency analysis can be found in \cite{diago2008a} (for the SMC) and \cite{diago2008b} (for the LMC). Many of the short-period variables have been found multiperiodic and some of them show the beating phenomenon due to the beat effect of close frequencies.

In Table~\ref{tab:1} we resume the percentages of short-period variable stars compared with the results obtained by 
\cite{gutierrez2007} for the Milky Way (MW).


\begin{table}
\caption{Percentages of short-period variables in the MC and in the MW.}
\label{tab:1}
\begin{tabular}{p{4cm}p{2cm}p{2cm}p{3cm}}
\hline\noalign{\smallskip}
			& MW	& LMC		& SMC		\\
\noalign{\smallskip}\svhline\noalign{\smallskip}
Pulsating B stars	& 16\%	& 6.9\%		& 4.9\%		\\
Pulsating Be stars	& 74\%	& 14.8\%	& 24.6\%	\\
\noalign{\smallskip}\svhline\noalign{\smallskip}
\end{tabular}
\end{table}

\section{Results}
\label{sec:3}

\begin{figure}[b]
\includegraphics[width=11.8cm]{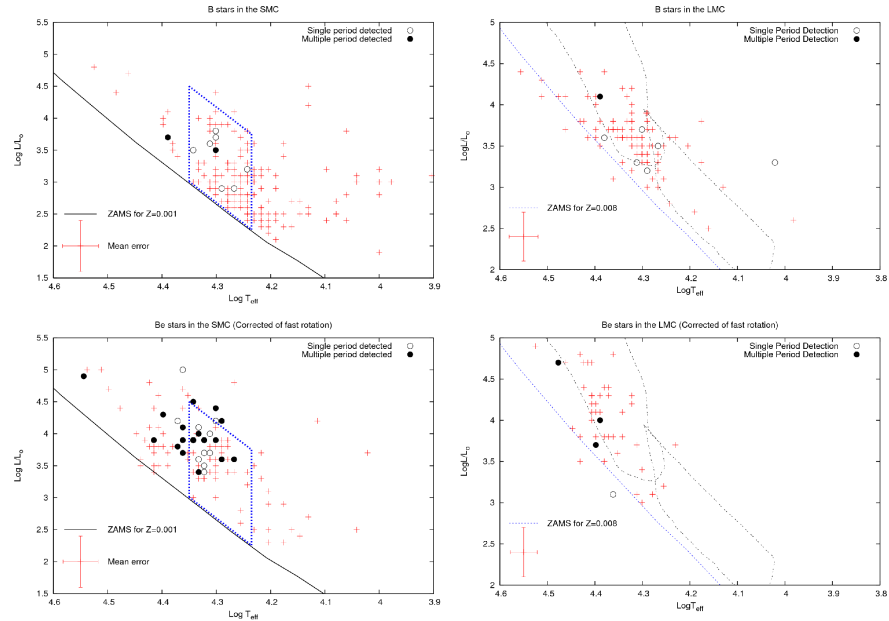}
\caption{Location of the B (top) and Be (bottom) star samples of the SMC (left panels) and the LMC (right panels) in the the HR diagram: single crosses represent stars in our sample, the empty circles represent single period detection and the filled ones multiple period detection. In the left panel (SMC), the dashed line delimits the suggested SPB instability strip for the SMC. In the right panel (LMC) the $\beta$ Cephei and the SPB boundaries at solar metallicities (taken from 
\cite{pamyatnykh1999}) are plotted only for reference.}
\label{fig:1}       
\end{figure}

\paragraph{\textbf{Small Magellanic Cloud}}
In Fig.~\ref{fig:1} we show the position in the HR diagram for the 9 short-period variable B stars found. All pulsating B stars are restricted to a narrow range of temperatures. Moreover, all stars but one have periods longer than 0.5 days, characteristic of SPB stars. Thus, we have suggested an observational SPB instability strip at the SMC metallicity, that it is shifted towards higher temperatures than in the Galaxy. We propose the hottest pulsating star in  our sample to be a $\beta$ Cephei variable. The reason is that it has two close periods in the range of p-mode galactic pulsators, but it also is the hottest pulsating star. If it is indeed a $\beta$ Cephei star, this would constitute an unexpected result, as the current stellar models do not predict p-mode pulsations at the SMC metallicities (see 
\cite{miglio2007}).

We have represented the 32 pulsating Be stars in the HR diagram (corrected of rapid rotation), and we have included the suggested SPB region at SMC metallicity. Most of the Be stars are located inside or very close to this region, suggesting that they are g-mode SPB-like pulsators. Three stars are significantly outside the strip towards higher temperatures, all of them multiperiodic, with periods lower than 0.3 days. Therefore, we propose that these stars may be $\beta$ Cephei-like pulsators.

The complete discussion of the results for the SMC is published in \cite{diago2008a}.

\paragraph{\textbf{Large Magellanic Cloud}}
In the LMC the search for variables is more difficult than in the SMC because they show more outbursts and irregular variations that prevent us from carrying the frequency analysis. In spite of this difficulty, we have obtain 7 short-period variables among the B star sample and 4 among the Be star sample. We have depicted in Fig.~\ref{fig:1} the short-period variables found in the LMC. As in the SMC, the hotter stars are those that are multiperiodic.

Our work with the results of the LMC is ongoing and will be published in 
\cite{diago2008b}.

%
%
%

\end{document}